\documentclass{aims_LENCOS}
\usepackage{amsmath}
  \usepackage{paralist}
  \usepackage{graphics} 
  \usepackage{epsfig} 
 \usepackage[colorlinks=true]{hyperref}
\hypersetup{urlcolor=blue, citecolor=red}

\def\currentvolume{X}
 \def\currentissue{X}
  \def\currentyear{200X}
   \def\currentmonth{XX}
    \def\ppages{X--XX}

\theoremstyle{definition}

\title[Logic operations demonstrated with]
      {Logic operations demonstrated with localized vibrations in a micromechanical cantilever array}

\author[M. Sato, N. Fujita and A. J. Sievers]{}


 \email{msato@kenroku.kanazawa-u.ac.jp}
 \email{sievers@ccmr.cornell.edu}

\thanks{}

\begin{document}
\maketitle

\centerline{\scshape M. Sato and N. Fujita }
\medskip
{\footnotesize
 \centerline{Graduate School of Natural Science and Technology, Kanazawa University}
   \centerline{ Ishikawa 920-1192, Japan}
} 

\medskip

\centerline{\scshape A. J. Sievers}
\medskip
{\footnotesize
 \centerline{ Laboratory of Atomic and Solid State Physics, Cornell University}
   \centerline{Ithaca, NY 14853-2501, USA}
}

\bigskip


\begin{abstract}
A method is presented for realizing logic operations in a micro-mechanical cantilever array based on the timed application of a lattice disturbance to control the properties of intrinsic localized modes (ILMs).  The application of a specific inhomogeneous field destroys a driver-locked ILM, while the same operation can create an ILM if initially no-ILM exists. Logic states ``1" and ``0" correspond to ``present" or ``absent" ILM.
\end{abstract}

\section{Introduction}

As a mechanical array decreases in size it becomes easier in the physical world for its vibrational dynamics to enter the nonlinear regime. One result for a micro or nano-mechanical array is that a localized nonlinear excitation called an intrinsic localized mode (ILM) can be generated.\cite{PRL,RMP} Some ideas\cite{Maniadis,Dick,Chen,Kenig,Kenig2} and applications\cite{spletzer,wiersig} for vibrational localization in small scale arrays have already been proposed. Experimental and simulation studies with micromechanical arrays have shown that in the nonlinear regime a steady state ILM can be produced by chirping the frequency of the driver\cite{Fajans} beyond the plane wave spectrum and then maintaining it there with a cw oscillator at constant amplitude.\cite{PRL} The resulting locked ILM is stable at a lattice site, and can be moved from one place to another by introducing a mobile impurity in the array.\cite{Opt} The interaction between an ILM and a vibrational impurity mode makes such motion possible. By using vibrational impurity mode dynamics a number of different processes have been uncovered such as seeding, annihilating, repelling and attracting driver-locked ILMs.\cite{RMP,ltp} When two vibrational bands are available dynamic local control is also possible by using a soliton in the acoustic-like  band to act on an ILM associated with the optic-like band.\cite{EL}  

Some characteristic ILM-local mode interaction properties that will be important for the current study are worth reviewing. For a nonlinear array with hard nonlinearity a stationary ILM can be created above the band states by using a chirped driver. When an impurity mode frequency is also above the band state frequencies but below the ILM frequency the ILM is attracted to it. (See Ref. \cite{RMP}, Fig. 9.) Two sets of frequency differences are important for the impurity mode control of the ILM. These are between the driver and the highest frequency pure-band state identified as $\Delta _{dm}=\omega _d-\omega _m$ and the frequency difference between the driver and the impurity mode called $\Delta _{di}=\omega _d-\omega _i$. Since the ILM is locked to the driver it is stable at the frequency shift $\Delta _{dm}$. When the ILM is near an impurity mode and $0<\Delta _{di}<\Delta _{dm}$, attraction occurs. In that case, the ILM releases some amplitude and becomes trapped at the impurity mode site. When the impurity mode is removed, the ILM amplitude is recovered. A sudden application of an impurity mode may end up un-locking the ILM from the driver and it will then disappear in an energy dissipation relaxation time; hence, to maintain the locked ILM its change in amplitude should occur over a longer time interval than the damping time.

By employing simulations it is demonstrated in this report that an inhomogeneous, harmonic force constant, time dependent perturbation applied to a nonlinear lattice can be used to control ILMs that are locked to a driver. It is shown that such manipulations can be used to produce logic operations. All logic operations, such as addition, subtraction, etc., can be made from basic sets of gates called complete sets. Complete sets are (i) NAND and (ii) NOR gates.\cite{Hurst} By application of the disturbance to the micromechanical cantilever array an inverter and NOR operation are demonstrated. 

In the next section the controlled lattice disturbance is described in some detail and inversion is demonstrated. Section 3 focuses on the NOR operation and the resulting truth table. Since signal processing become easier if more processes are available additional operations are presented in Section 4.  Observed transient phenomena are discussed in the remaining section followed by a summary.

\begin{figure}[htp]
\begin{center}
  \includegraphics[width=4in]{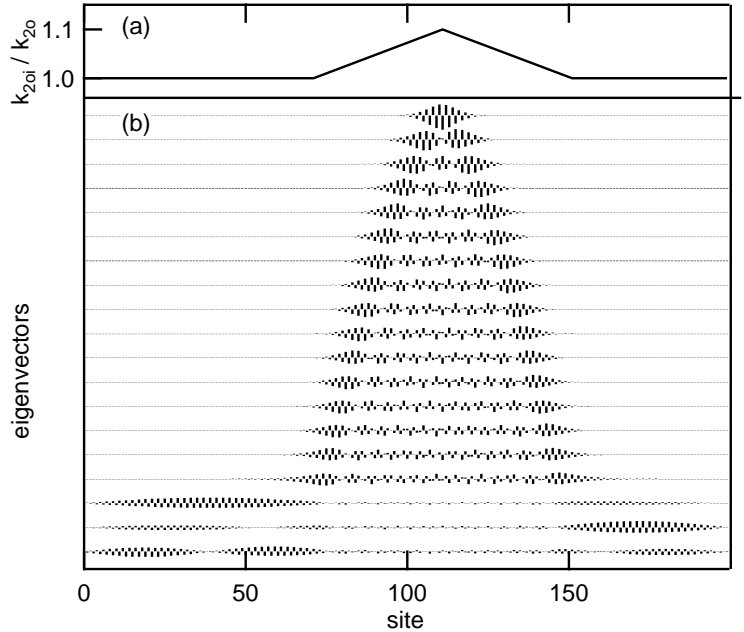}\\
  \caption{(a) Spatial inhomogenous pattern applied to manipulate ILMs. The disturbance is characterized as the ratio of the on-site impurity spring constant to the pure one. (b) Linear eigenvectors from the highest resonance frequency down to the 19-th mode. The top 16 modes show a variety of localization behavior. Extended band modes begin at the 17-th mode. Total numbers of modes = 200.}\label{fig1}
  \end{center}
\end{figure}

\section{Inversion}

Our micro-mechanical array simulation model is based on the experimental observations for cantilever arrays described in Ref. \cite{RMP}. It makes use of the lumped element model equation for cantilever $i$:

\begin{equation}\label{1}
 \begin{split}
 &m_i \ddot x_i  + \frac{{m_i }}{\tau }\dot x_i  + k_{2Oi} x_i  + k_{4O} x_i^3  \\ 
 &+ \sum\limits_j {k_{2I}^{(j)} (2x_i  - x_{i + j}  - x_{i - j} )}  \\ 
 &+ k_{4I} [(x_i  - x_{i + 1} )^3  + (x_i  - x_{i - 1} )^3 ] = m_i \alpha \cos \omega _d t \\ 
 \end{split} 
\end{equation}
where  $m_i$  is the mass,  $\tau$ is a life time,  $k_{2Oi}$ and $k_{4O}$ are onsite harmonic and quartic spring constants, $k_{2I}^{(j)}$ is a harmonic intersite spring constant of $j$-th nearest neighbor, $k_{4I}$ is a quartic intersite spring constant, $\alpha $ is an acceleration, and $\omega _d$ is the driver frequency. The array is made from an alternative sequence of long and short cantilevers to insure coupling to the optic branch with the uniform acceleration force provided by a PZT transducer. Thus,  $m_i$ and $k_{2Oi}$ are alternatively repeated along the array. The nonlinear components $k_{4O}$ and $k_{4I}$ are both positive and $k_{4I}>>k_{4O}$. There are a total of 200 modes. The ILM is generated above the highest linear resonant band frequency (137.1 kHz). The center of the ILM is at the short cantilever site (odd number site in the simulation). The driver frequency and amplitude are fixed $\omega _d/2\pi =139 $ kHz and $\alpha =500 $ m/s$^2$ throughout the paper.

The lattice disturbance is introduced as a time and spatial dependence of the onsite harmonic spring constant over a fixed region. The spatial pattern to be applied to the array is shown in Fig. \ref{fig1}(a). It extends over a third of the lattice and the maximum increase in the onsite harmonic spring constant is 10\%. As shown in Fig. \ref{fig1}(b) at the time of maximum application the top 16 modes range from extended island modes at low frequencies to local modes at high frequencies. 

\begin{figure}[htp]
\begin{center}
  \includegraphics[width=4in]{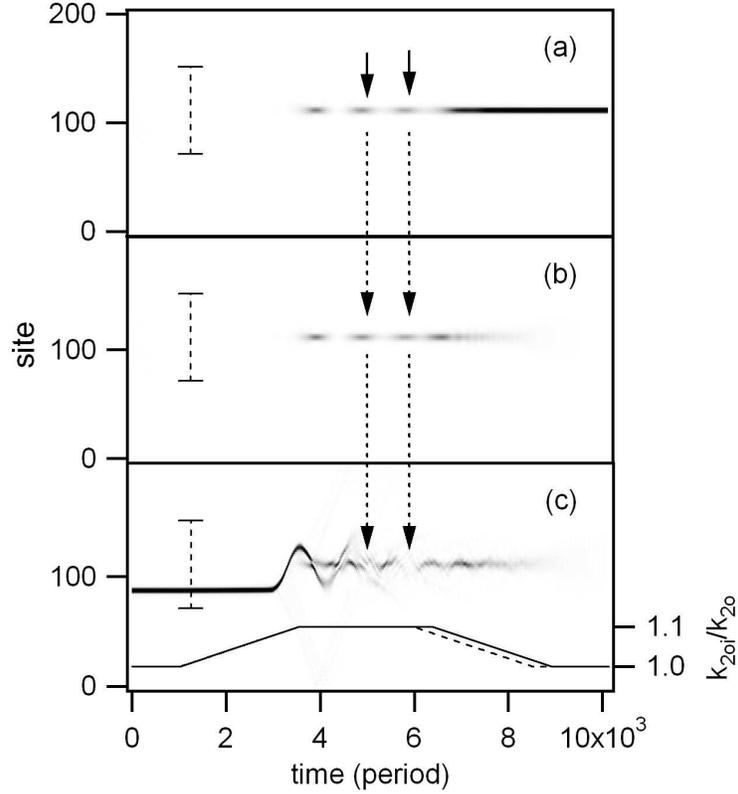}\\
  \caption{Simulated inverter logic in a micromechanical cantilever array. Dark region corresponds to high energy. The horizontal lines separated by a vertical dashed line identify the boundaries of the triangular disturbance. (a) One ILM is created from the no ILM state. (b) If the disturbance is removed a half AM period earlier than (a), see dashed line in (c), no ILM remains. (c) ILM at site 87, shifted by 24 sites from the maximum, is destroyed by the lattice perturbation. The solid line shows the time dependence of disturbance used in (a) and (c). Vertical arrows indicate different times where energy has maxima in (a).}\label{fig2}
  \end{center}
\end{figure}

Figure \ref{fig2}(a) shows a simulation result for the time dependence of a locked ILM in the presence of the time-dependent spatial disturbance. The driver frequency is fixed at 139.0 kHz, so $\Delta _{dm}=1.9$ kHz while $\Delta _{di}=0.05 $ kHz at the disturbance maximum. After turning the disturbance on instantaneously a localized pulsing structure appears at its center, indicating an amplitude modulation (AM) of the trapped ILM. Upon reducing the strength of the perturbation to 0 in a time characterized by the ILM relaxation time,  $\tau $, one ILM remains in the system. The presence or absence of the ILM depends crucially on the phase of its AM modulation at the time of the removal of the disturbance. Fig. \ref{fig2}(b) shows that if the removal of the disturbance is one half period AM earlier than for frame (a), no ILM remains.  At least three cycles of  ``presence" and ``absence" of ILMs are seen in simulations when the removal time is scanned with respect to the AM period. The difference in the end results, i.e., ILM or no ILM, is observed in the time region where the pulsing structure is clearly seen.  

Figure \ref{fig2}(c) shows an ILM at site 87 interacting with a disturbance whose growth and removal time is the same, $\sim \tau $. In this case the initial ILM is attracted to island modes of larger and larger amplitude shifting it towards the local mode at center of the disturbance (site 111). Notice the interesting result that the ILM crosses the disturbance center due to its translational inertia and oscillates about that ``equilibrium" position.  At the same time, the AM of the nonlinear mode at the disturbance center begins to grow in amplitude. This nonlinear mode is an incipient ILM that stabilizes once the disturbance is removed, similar to that shown in Fig. \ref{fig2}(a). At a later time (around $5000\sim 7000$ periods) the phase of the AM is shifted by 180 degrees compared with its value in Fig. \ref{fig2}(a). Since the phase is inverted, the end result is also inverted even if the removal time interval of the disturbance is the same as frame (a). (If the temporal pattern in (b) is used, the result is an ILM.) Thus, the same disturbance operation creates an ILM if there is none initially as in (a), or destroys the ILM if one initially exists as in frame (c). It should be noted that the combination of the translation and the AM with respect to the time disturbance causes the results shown in Fig. \ref{fig2}(c) to depend to some extent on the initial relative location of the ILM.

The first two frames in Fig. \ref{fig2} have also been obtained in another set of simulations where the time dependent pattern shown at the bottom of Fig.  \ref{fig2}(c) is used. In this case the time dependence of the disturbance is the same in frames (a) and (c) with its finish shifted by 1/2 period in (b). As expected the AM patterns are the same between $t = 4000\sim 6000$ periods in (a) and (b), while the pattern is inverted in (c) as illustrated by the vertical arrows shown in all three frames. The 180 degree phase shift of the amplitude modulation in (c) causes the end result to be reversed from that in (a). It is the presence and location of the initial ILM that changes the result. Assigning the existence and absence of ILMs as ``1" and ``0" the simulations described here demonstrate inverter action in the array.

We have investigated the distance dependence of the ILM from the disturbance center.  Table \ref{table1} summarizes the results.  The third row of Table \ref{table1} gives the initial ILM location dependence for the inverter action. A ``0" indicates inverter action because the disturbance creates one ILM in the absence of an initial ILM. On the other hand, ``1" indicates ``logical 1" because existence or no-existence of the initial ILM doesn't matter.  When the initial location is too far removed, there is no reaction. When the location is too close to the disturbance, the ILM remains at the disturbance center. For the middle distance, ``1" and ``0" appear periodically as the distance is increased from the center and the period decreases with the distance.

\begin{table}[htp]
\begin{center}
\caption{Tabulated logic operations for the spatial dependences of single and dual initial ILMs. Row 1:  Initial ILM position. Row 2: Distance from the center of the disturbance. Row 3: Dependence for the temporal pattern in Fig. 2(c) solid curve. ILMs are stable only on the short cantilever sites (odd number). A ``0" or ``1" means absence or existence of the ILM after the disturbance application. ``x" means ``no reaction with the disturbance". Row 4: For the dual input case, the other ILM is positioned at the opposite symmetric cantilever position. If two ILMs are too close to each other the ILMs interact and those cases are indicated by ``xx".  Row 5: EN, ND, and NR stand for EXOR-NOT( $\overline {\bar AB + A\bar B} $), NAND($\overline {AB} $ ) and NOR ($\bar A\bar B = \overline {A + B} $ ) logic operations. The ``-" at site 93 stands for non-symmetric action of the symmetric site.}
\label{table1}

\begin{tabular}{c|c|c|c|c|c|c|c|c|cc}
\cline{1-9}
\multicolumn{1}{|c|}{initial location} & 75 & 77 & 79 & 81 & 83 & 85 & 87 & 89 &  &  \\
\cline{1-9}
\multicolumn{1}{|c|}{distance} & 36 & 34 & 32 & 30 & 28 & 26 & 24 & 22 &  &  \\
\cline{1-9}
\multicolumn{1}{|c|}{single input} & x & x & 0 & 1 & 0 & 1 & 0 & 1 &  &  \\
\cline{1-9}
\multicolumn{1}{|c|}{dual input} &  &  & 1 & 0 & 0 & 0 & 0 & 0 &  &  \\
\cline{1-9}
\multicolumn{1}{|c|}{logic} &  &  & EN & ND & NR & ND & NR & ND &  &  \\
\cline{1-9}
\multicolumn{1}{c}{} & \multicolumn{1}{c}{} & \multicolumn{1}{c}{} & \multicolumn{1}{c}{} & \multicolumn{1}{c}{} & \multicolumn{1}{c}{} & \multicolumn{1}{c}{} & \multicolumn{1}{c}{} & \multicolumn{1}{c}{} & \multicolumn{1}{c}{} &  \\[-6pt]
\cline{2-11}
 & 91 & 93 & 95 & 97 & 99 & 101 & 103 & 105 & \multicolumn{1}{|c|}{107} & \multicolumn{1}{c|}{109} \\
\cline{2-11}
 & 20 & 18 & 16 & 14 & 12 & 10 & 8 & 6 & \multicolumn{1}{|c|}{4} & \multicolumn{1}{c|}{2} \\
\cline{2-11}
 & 0 & 1 & 0 & 1 & 1 & 0 & 1 & 1 & \multicolumn{1}{|c|}{1} & \multicolumn{1}{c|}{1} \\
\cline{2-11}
 & 0 & 0 & 0 & 0 & 0 & 0 & 1 & 0 & \multicolumn{1}{|c|}{xx} & \multicolumn{1}{c|}{xx} \\
\cline{2-11}
 & NR & - & NR & ND & ND & NR &  & ND & \multicolumn{1}{|c|}{} & \multicolumn{1}{c|}{} \\
\cline{2-11}
\end{tabular}

\end{center}
\end{table}

\begin{figure}[htp]
\begin{center}
  \includegraphics[width=4in]{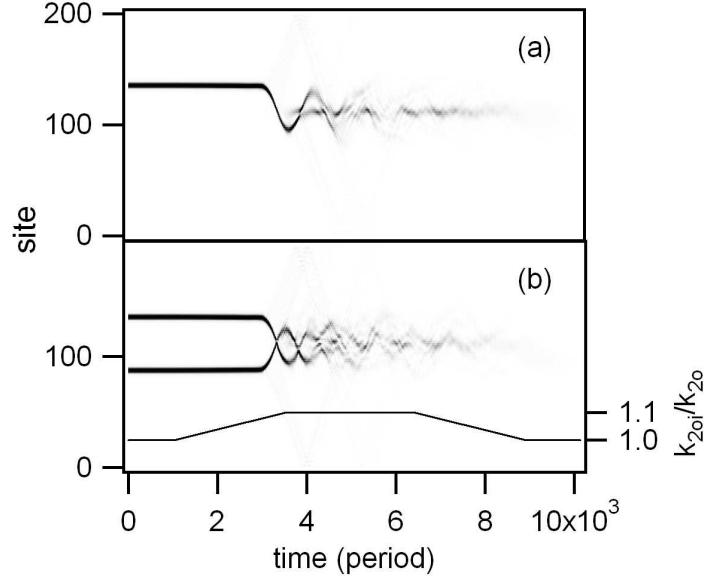}\\
  \caption{NOR logic operation.  (a) Initial ILM at site 135, shifted by 24 sites to the other side of the disturbance maximum as applied in Fig.  \ref{fig2}(c). (b) Two simultaneous ILMs at site 87 and site 135 end in no ILM after the perturbation is applied and removed.}\label{fig3}
  \end{center}
\end{figure}

\section{NOR operation}
Figure  \ref{fig3}(a) shows that when the initial ILM is located on the symmetrically opposite side of the disturbance, at site 135, and the same time dependence of Fig. \ref{fig2}(c) is applied, the AM phase of the ILM is inverted. [Compare Fig. \ref{fig2}(c) with Fig. \ref{fig3}(a).] When two ILMs are placed at sites 87 and 135 then at the end of the perturbation cycle no ILM remains as shown in Fig. \ref{fig3}(b).  In this case, the smooth AM observed in Fig. \ref{fig2}(a) is more difficult to see.

\begin{table}[htp]
\begin{center}
\caption{Truth table of the logic operation. A and B are inputs at sites 135 and 87. Output site is 111. ``1" and ``0" mean existence and absence of ILMs. From this table, we can see that the logic operation  $\bar A\bar B = \overline {\left( {A + B} \right)}$ is realized. It corresponds to the NOR.}
\label{table2}
\begin{tabular}{|c|c|c|c|}
\hline
A(135) & B(87) & Output(111) & Figure \\
\hline
0 & 0 & 1 & \ref{fig2}(a) \\
\hline
0 & 1 & 0 & \ref{fig2}(c) \\
\hline
1 & 0 & 0 & \ref{fig3}(a) \\
\hline
1 & 1 & 0 & \ref{fig3}(b) \\
\hline
\end{tabular}
\end{center}
\end{table}

A truth table for the triangular disturbance operation is presented in Table \ref{table2}.  Inputs are the initial existence or absence of ILMs at sites 87 and 135. The output is site 111, the peak of the impurity pattern in Fig. \ref{fig1}.  The results from Figs. \ref{fig2} and \ref{fig3} are summarized in Table \ref{table2}. Output ``1" is obtained for input ``00".  The operation is expressed as  $\bar A\bar B = \overline {\left( {A + B} \right)} $ and is a NOR.

If the logic ``1" results for two ILMs placed at other symmetric distances with respect to the disturbance center then logic operations are still possible as long as the ``1" occurs with high reliability.  In this case, the 3rd column, last row of the truth table will be changed to ``1" and the operation expression will be$\bar A\bar B + AB$, that is, an EXOR-NOT gate. In Table \ref{table1}, 4th row, the results for two symmetric inputs (dual input) are summarized. When a single input ends in ``0", the operations are classified into NOR or EXOR-NOT depending on the dual input results as mentioned above. For such operations observing a well-defined amplitude modulation for the ILM is the most important feature. On the other hand, there are cases of single input ending in ``1" and dual inputs results ending in "0" in Table \ref{table1}. Such operation can be expressed as $\bar A\bar B + A\bar B + \bar AB = \overline {AB}$, i.e., a NAND gate. In addition, there is a slight non-symmetric behavior at site 93 where the opposite site acts as the inverter site. Such a non-symmetric property is probably due to the non-symmetric location of the disturbance with the fixed boundary condition. However, such a non-ideal property may disappear with a larger lattice.

\section{Other operations}
\begin{figure}[htp]
\begin{center}
  \includegraphics[width=4in]{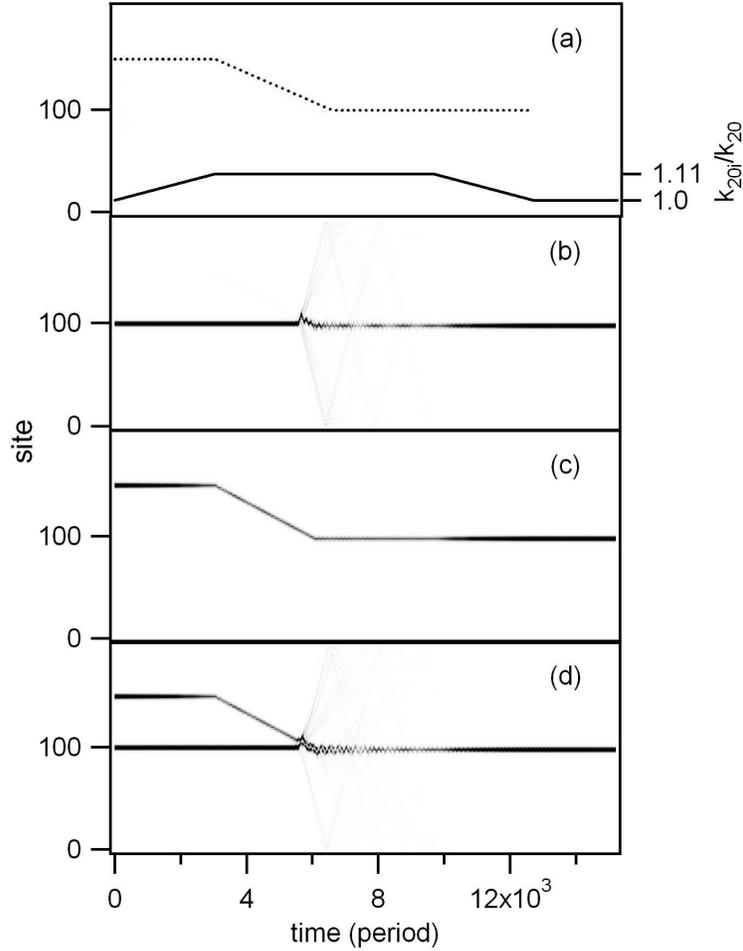}\\
  \caption{OR operation. One triangular disturbance is translated from site 149 to 99 with time as illustrated by the dashed line in panel (a). The time dependence of the disturbance itself is shown by the solid line at the lower part of panel (a). The width of the disturbance is the same as for the previous figure. (a) With the no ILM initial condition, nothing appears at output site 99.  (b) Starting with one ILM at site 99. (c) Starting with one ILM at site 149. (d) Starting with two ILMs at site 99 and 149. Operations shown here indicate OR logic, that is, it matches with Boolean equation  $A+B$ where $A$ and  $B$ are inputs at site 99 and 149.}\label{fig4}
  \end{center}
\end{figure}

With a larger variety of possible operations available signal processing can become much easier. Figure \ref{fig4} shows an OR operation, again produced with the triangular disturbance.  Input sites are 99 and 149. One triangular disturbance is moved along the dashed line as a function of time. The disturbance is sufficiently weak so that it doesn't create an ILM by its application if there is no ILM present initially. Figures \ref{fig4}(b) and (c) are for the one ILM present initial condition. The result is the same as the manipulation by a single impurity mode. Figure \ref{fig4}(d) starts with two ILMs and they collide at site 99. In most cases the collision of two locked ILMs by the manipulation of a single ILM ends with no ILM. However, for the triangular disturbance, the result is one ILM.

\begin{figure}[htp]
\begin{center}
  \includegraphics[width=4in]{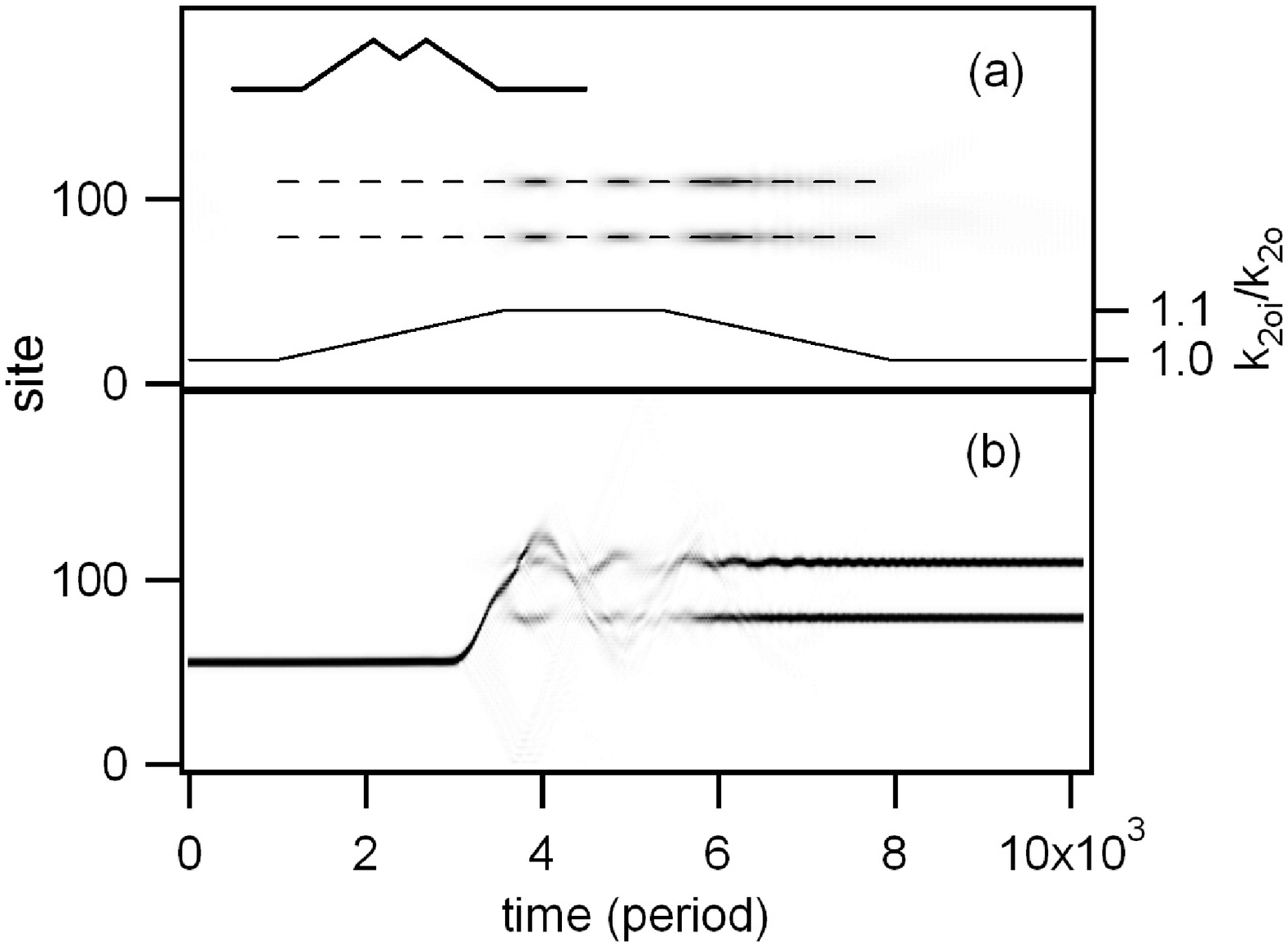}\\
  \caption{Duplication of an ILM. Two triangular disturbances are used centered at sites 79 and 109 with the same width as in the previous figures. Two dashed lines indicate their positions. These disturbances overlap in the middle as illustrated with the pattern shown in the left-top corner. Time dependence of the disturbance itself is shown in lower part of (a). (a) With no initial ILM, no ILMs appear at site 79 and 109. (b) With an initial ILM at site 55, two ILMs appear as outputs after the application of the disturbance.}\label{fig5}
  \end{center}
\end{figure}

Other useful operations are duplication and swapping. Often one signal is used many times in general logic circuits. The signal is simply branched in such circuit diagrams; however, in the nonlinear array for the ILM operation, each ILM signal must be prepared before the operation. Thus, duplication of an ILM is required. Figure \ref{fig5} shows such a process for two triangular disturbances that slightly overlap. In this case two ILMs appear although only one is initially present.

\begin{figure}[htp]
\begin{center}
  \includegraphics[width=4in]{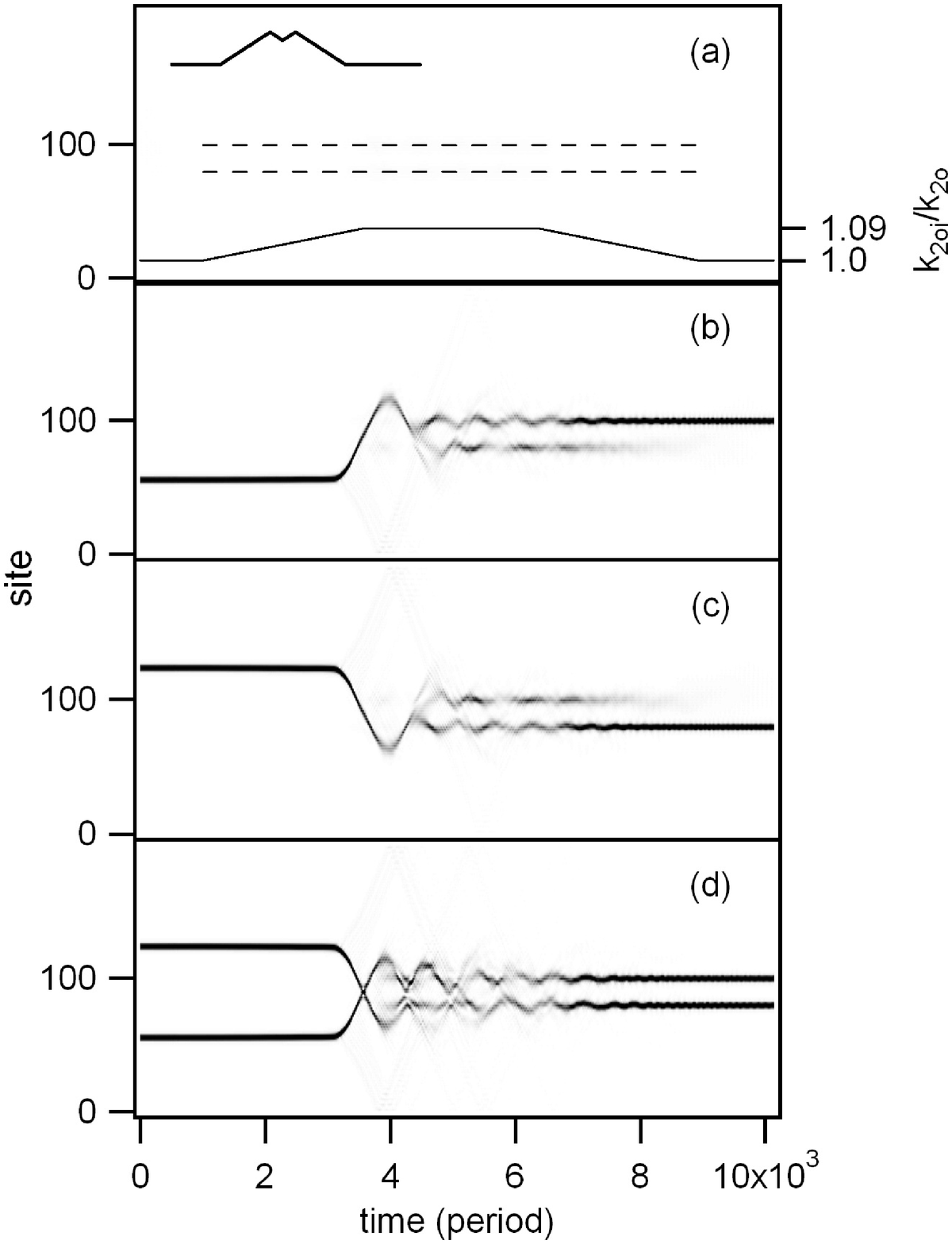}\\
  \caption{Swapping of ILMs. Two triangular disturbances are used centered at sites 79 and 99 (positions indicated by two dashed lines) with the same width as in the previous figure. These disturbances overlap in the middle as illustrated with the pattern shown in the left-top corner. Time dependence of the disturbance is shown in lower part of (a). (a) No initial ILM results in no outputs at site 79 and 99. (b) An initial ILM at sites 55 moves to site 99. (c) An initial ILM at site 123 moves to site 79. (d) Starting with two ILMs, the end result is two ILMs at sites 99 and 79. }\label{fig6}
  \end{center}
\end{figure}

Branches are often seen in circuit diagrams and the crossing of circuit lines may exist at many locations. No interaction occurs between two crossing lines. For the one-dimensional nonlinear array, this corresponds to moving one ILM signal from a site to the other side of another ILM signal position. Figure \ref{fig6} shows such a case. The ILM signal at site 55 propagates to site 99, and that initially on site 123 is transmitted to site 79.

\section{Discussion}
Two transient phenomena have been observed in the production or destruction of an ILM. (1) Oscillation of the frequency locked ILM around the point of maximum disturbance and (2) transient amplitude modulation.\cite{Chaos} The first signature indicates that the ILM has translational inertial while the source of the second effect is more subtle.

For a linear oscillator the AM frequency is the difference frequency between the driver and the oscillator, i.e., $\Delta _{di}=0.05 $kHz, but the observed AM oscillation frequency in the cases presented here is $\sim 0.14$ kHz. In addition, the lifetime of the AM ($\sim 2300$ periods) is longer than the damping time, $\tau \sim 1200$ periods. Thus, the observed AM cannot be explained by a simple linear calculation, instead it must be strongly influenced by the nonlinear effect.  

Such AM has already been observed in the sideband spectrum of driver locked ILMs\cite{Chaos} and has been identified with a low frequency deformation mode of the ILM itself.\cite{AM} The oscillation parametrically receives energy from the main driver so this parasitic oscillation has a longer lifetime than for the linear case. The added translational component observed here may be the source of the large AM depth.

Bistability of a driver locked ILM is a known property\cite{Rossler} and such bistability is also known to occur for a Duffing oscillator.\cite{Nayfeh} By using the ILM eigenvector and a transformation method described in Ref. \cite{daSilva} and integrating along the array (not along the cantilever), a Duffing model has been created with effectively the same parameters as used in the above simulations. Similar to the array case, locked and unlocked final states appear alternatively when the perturbation removal time is changed by 1/2 the AM period. Thus, one nonlinear oscillator with a fixed driver is sufficient to have both amplitude modulation and two end results, namely, locked and un-locked states.

Even if amplitude oscillation and bistability are explained by the single nonlinear oscillator model, the multiple input operation demonstrated here requires many degrees of freedom. Since the end result is sensitive to the phase of the AM of the ILM, logic operations are possible as long as a method to modify the AM phase can be found. 

The triangular disturbance has a definite advantage for controlling ILMs. Since the locked ILM has fixed excitation amplitude, interaction of two nearby ILMs usually causes repulsion interaction between them. The strong repulsion often ends up producing the no ILM result. Although a single impurity mode may also manipulate ILMs, the triangular disturbance has an advantage in terms of multiple ILM interactions because interactions will occur over a larger region of space. 

One question is how to make the operations shown here work for a large 1-D array. In this case, multiple ILM interactions are possible. Ideally, there should be no interaction during the crossing; however, locked ILMs are stationary stable and strongly interact when they cross each other. More desirable would be a localized excitation with a weaker interaction during a rapid crossing so that swapping would be easier. Because solitons do not interact with each other and because solitons run while ILMs do not, converting an ILM into a soliton and then after the swap converting the soliton back to an ILM, would make the swapping operation highly reliability in a large array.  Just exactly how to engineer such a process is not yet clear although the results in Ref. \cite{EL} may provide clues. So far it appears that handling the ILM-ILM interaction between stationary stable ILM is much simpler than dealing with the still to be explored ILM-soliton interaction.

\section{Summary}
Logic operations inverter and NOR have been demonstrated in simulations involving a cantilever array. The triangular spatial perturbation to the array vibrational dynamics operated off-center to the initial ILM state inverts the existence or absence of the ILM. The operation is explained by a phase shift of the AM produced by the attraction of the initial ILM(s) and the growing (or decaying) array of impurity modes produced by the time dependent perturbation.  Any kind of logic operation can be achieved by combinations of the inverter and NOR.

The operations shown here are based on the transient response AM together with the bistability of this Duffing-like oscillator. Such a process also would be expected to occur in other kind of MEMS oscillator arrays. Finally this proof of principle study of possible logic operations may apply to actuator arrays and sensor arrays. If and when lattice dynamics ILMs in condensed matter systems can be manipulated then information processing may also be possible.

\section*{Acknowledgements}
This work was supported in part by NSF-DMR Grant No. 0906491, DOE No. DE-FG02-04ER46154 and by JSPS-Grant-in-Aid for Scientific Research No. (B) 18340086.

\medskip
Received xxxx 20xx; revised xxxx 20xx.
\medskip

\end{document}